\documentclass[11pt,twoside]{article}
\usepackage{jltp}
\usepackage{graphics}
\title{Is There a Unified Description of Conductivity of Layered Cuprates ?\cite{Q}}

\author{C. C. Almasan, E. Cimpoiasu, G. A. Levin,\address{Department of
Physics, Kent State
University, Kent, OH 44242, USA} H. Zheng$^{*}$, A. P. Paulikas$^{*}$, and B.
W. Veal\address{Materials Science Division,
Argonne National Laboratory, Argonne, IL 60439, USA}}

\runninghead{C. C. Almasan {\it et al.}}{Is There a Unified Description of
Conductivity}

\begin{document}

\begin{abstract}
We present a novel  approach to  the analysis of   the normal state in-plane
$\sigma_{ab}$ and out-of-plane
$\sigma_{c}$ conductivities of  anisotropic layered crystals such as oxygen deficient
$YBa_{2}Cu_{3}O_{x}$.
It can be shown that the resistive anisotropy is determined by the ratio
of the phase coherence  lengths in the respective directions; i.e.,
$\sigma_{ab}/\sigma_c=\ell_{ab}^2/\ell_c^2$.
From the  idea that at all doping levels and temperatures $T$
the out-of-plane transport  in these crystals is incoherent,
follows that $\ell_c$ is T-independent, equal to the spacing $\ell_0$
between the
neighboring bilayers.  Thus, the T-dependence of $\ell_{ab}$ is given by
the measured anisotropy,  and  $\sigma_{ab}(\ell_{ab} )$ dependence is obtained by
plotting $\sigma_{ab}$ vs $\ell=(\sigma_{ab}/\sigma_c)^{1/2}\ell_0$.  The analysis
of several single
crystals of $YBa_{2}Cu_{3}O_{x}$  ($6.35<x<6.93$)
shows that for all of them $\sigma_{ab}(\ell )$ is described by a universal
dependence
$\sigma_{ab}/\bar\sigma =f(\ell/\bar\ell )$ with doping dependent parameters
$\bar\sigma$ and $\bar\ell$.

\end{abstract}

\maketitle

\vspace{0.3in}

The temperature $T$ dependence of the  in-plane $\rho_{ab}$ and out-of-plane
$\rho_c$ normal state
resistivities  of layered  cuprates such as  $YBa_{2}Cu_{3}O_{7-\delta}$
shows  a diverse behavior ranging from metallic  in optimally-doped and
overdoped samples,
to  the coexistence of  metallic $\rho_{ab}$ and nonmetallic $\rho_c$
in moderately underdoped crystals, to insulating
$\rho_{ab}$ and $\rho_c$ in
strongly underdoped specimens.  A common feature  of all cuprates is a very
large ($\sim 10^2-10^5$) and
temperature dependent  anisotropy  $\rho_c/\rho_{ab}$.  Quasiclassical
estimates of the mean free path in the c-direction
indicate that  it is less than the interlayer spacing even in the least
anisotropic systems such as optimally-doped
$YBa_{2}Cu_{3}O_{7-\delta}$\cite{Cooper}.
This means that  the
normal-state transport in the c-direction is incoherent  at all
temperatures so that the
wave function of the electrons loses its coherence over the shortest
possible distance, the interlayer spacing $\ell_0$, while  the
coherence length [hereafter called Thouless length (TL)] in the $CuO_2$ planes  is much larger
than the size of the unit cell, as
evidenced by the values of $\rho_{ab}$ which are well below the Mott's
limit\cite{Cooper}.   

The layered crystals with incoherent interlayer transitions represent a
unique system
in which the strong interlayer decoherence of the charge carrier's wave
function could be the reason for the strong $T$ dependence of the
resistive anisotropy.
This follows from the general relationship between conductivities\cite{PRL}:
\begin{equation}
\frac{\sigma_{ab}}{\sigma_{c}}=\frac{\ell^2}{\ell_c^2}.
\end{equation}
Here  $\ell$ is the in-plane TL and $\ell_c$ is the out-of-plane TL.
In conventional metals, the two phase coherence lengths change with temperature
at the same rate
so that  $\sigma_{ab}/\sigma_{c}= const$.  The strong interlayer
decoherence in layered crystals
(of still unknown origin) implies that $\ell_c$ is
temperature independent, equal to the interlayer spacing; i.e., $\ell_c=\ell_0$.
As a result, the anisotropy is temperature dependent, reflecting  the
$T$ dependence of the  in-plane  phase coherence length.
Thus,  the layered crystals with incoherent interlayer transitions offer
the opportunity
to obtain experimentally both the in-plane TL (by measuring $\rho_c/\rho_{ab}$) and  the
functional dependence of
$\sigma_{ab}$ on the phase coherence length [by
plotting $\sigma_{ab}$ vs $ (\rho_{c}/\rho_{ab})^{1/2}\ell_0$].

We performed measurements of both 
$\rho_{ab}$ and  $\rho_{c}$ as a function of
T ($ T_c<T\le 300\;K $) on $YBa_2Cu_3O_{x}$ ($6.35 \leq x \leq 6.93$) single
crystals using the four-point method,  as well as the multiterminal technique
(flux-transformer geometry\cite{Jiang}).
Figure 1 is a plot of $\sigma_{ab}$  vs 
$\ell =(\rho_c/\rho_{ab})^{1/2}\ell_0$ with 
$\ell_0=11.7\AA$. The open  symbols represent the raw data.
An idea  outlined in Ref. 3
is that  the variation with doping of the number of carriers and the amount
of disorder  does not
alter fundamentally the
$\sigma (\ell )$ dependence, and that the effect  of these changes can be
incorporated into
two constants, one  of which, $\bar\sigma $,  normalizes the magnitude of
the in-plane conductivity,
and the other, $\bar\ell $,   normalizes the in-plane Thouless length, namely,
\begin{eqnarray}
\frac{\sigma_{ab}}{\bar\sigma }=f\left (\frac{\ell}{\bar\ell}\right
);\;\;\;f(1)=1.
\end{eqnarray}
Here $f(y)$ is the same function
for a given class of single crystals, such as $YBa_{2}Cu_{3}O_{x}$.
Varying oxygen content  in this case  only changes the values of the
normalization constants
$\bar\sigma$ and $ \bar\ell$, shifting the respective segments  of
$\sigma_{ab}(\ell )$
along a  common trajectory.
\begin{figure}[htbp]
\vspace*{-0.14in}             
\hspace*{0.7in}
\includegraphics*{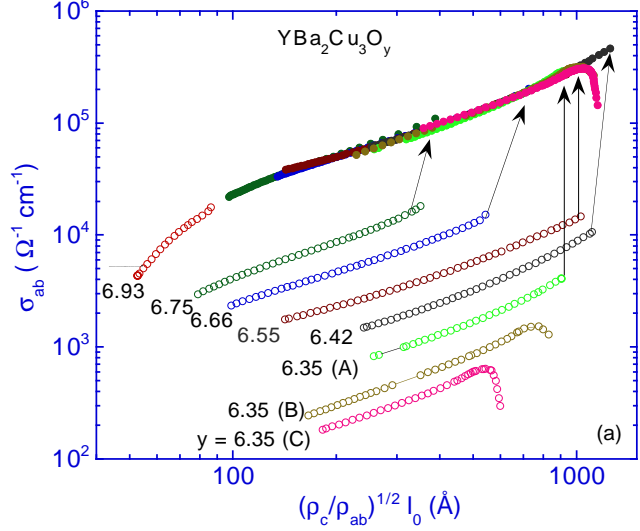}
\caption{\it In-plane conductivity $\sigma_{ab}$  of $YBa_2Cu_3O_{x}$
single crystals plotted vs phase coherence
length defined as $\ell= (\rho_c/\rho_{ab})^{1/2}\ell_0$, with
$\ell_0=11.7\AA$.
The open symbols correspond to the raw data.  For all samples
the lowest anisotropy corresponds to highest temperature and vice versa.
The filled symbols which form a
trajectory [Eq. (2)] are obtained by shifting the open symbol
segments parallel to themselves
as indicated by the arrows.  
}
\end{figure}

The continuous curve indicated in Fig. 1 by the filled
symbols is generated by shifting the segments corresponding to different oxygen
concentrations parallel to themselves (as indicated
by the arrows),   matching the values and the slopes of the {\it overlapping}
segments. The data for  the $x=6.93$ sample  were   used as reference.
On a log-log plot, such  shifts are equivalent to  change of  normalization
constants $\bar\sigma$ and $\bar\ell$.    Through this
procedure we obtained the  trajectory
$\sigma_{ab}(\ell )$ given by Eq. (2), with the values of
$\bar\sigma $ and $\bar\ell $ corresponding to  the oxygen content
of the reference sample ($x=6.93$). The branching point reflects 
the metal-insulator transition.
\begin{figure}[htbp]
\vspace*{0.1truein}             
\hspace*{0in}
\resizebox{\textwidth}{!}{\includegraphics*{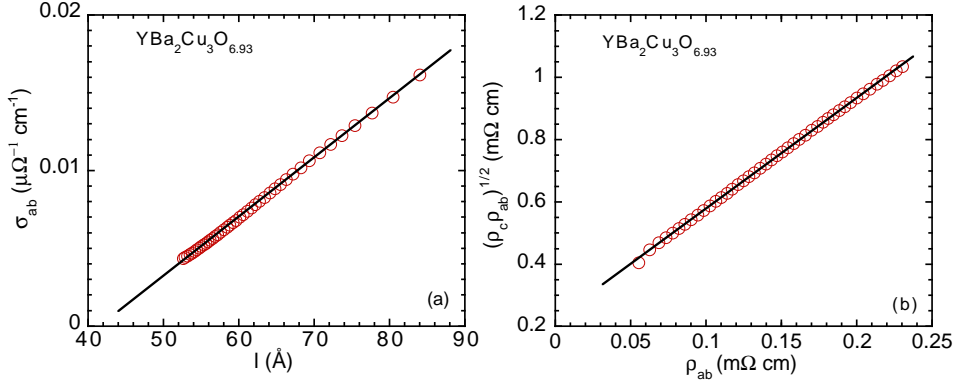}}
\caption{\it (a) In-plane conductivity of $YBa_2Cu_3O_{6.93}$ crystal plotted vs phase coherence length  
$\ell$; (b) same data plotted as 
$(\rho_{c}\rho_{ab})^{1/2}$ vs $\rho_{ab}$  [see Eq. (5)].
}
\end{figure}

Here we consider in more detail the functional dependence
$\sigma_{ab}(\ell )$
of the optimally doped $YBa_2Cu_3O_{6.93}$ single crystal.
Figure 2(a) shows its conductivity plotted vs $\ell $.
This $\sigma_{ab}(\ell)$ dependence is well described by:
\begin{equation}
\sigma_{ab}= q (\ell -\xi );\;\; \ell> \xi .
\end{equation}
Based on Eq. (1), the corresponding $\sigma_c$ is given by
\begin{equation}
\sigma_{c}= q\ell_0^2 \left (\frac{1}{\ell} - \frac{\xi }{\ell^2}\right  ).
\end{equation}
Note that $\sigma_{c}$ is metallic ($d\sigma_{c}/d\ell >0$)
for $\ell < 2\xi $, and nonmetallic ($d\sigma_{c}/d\ell <0$) for $\ell > 2\xi $.  Since the
phase coherence length
monotonically increases with decreasing temperature,  $\sigma_{c}$ is
metallic at high $T$,
reaches a maximum when $\ell (T) = 2\xi $, and decreases with further decreasing
$T$, while $\sigma_{ab}$
remains  metallic at all $T$. Equations (3) and (4) are equivalent to
the following relationship
between resistivities:
\begin{equation}
(\rho_{c}\rho_{ab})^{1/2}= \bar\rho + \left (\frac{\xi}{\ell_0}\right
)\rho_{ab};\;\; 
\bar\rho=\frac{1}{q\ell_0}.
\end{equation}
Figure 2(b) is a plot of $(\rho_{c}\rho_{ab})^{1/2}$ vs $\rho_{ab}$. A fit of the data with Eq.
(5) gives  $\bar\rho \approx 0.23\;m\Omega\;cm $ and $\xi \approx 41\;\AA$. Thus, the
temperature dependence of
$\rho_{c}$ is determined by that of $\rho_{ab}$:
\begin{equation}
\rho_{c}=   \left (\frac{\xi}{\ell_0}\right )^2\rho_{ab}  +2\left
(\frac{\xi}{\ell_0}\right )\bar\rho
+ \frac{\bar\rho^2}{\rho_{ab}}.
\end{equation}
Specifically, when $\rho_{ab}=\alpha_{ab} T$, which is characteristic of
the optimally doped $YBa_2Cu_3O_{7-\delta}$,
\begin{equation}
\rho_{c}=  \beta_c + \alpha_c T +\frac{\gamma_c}{T}.
\end{equation}
Note, that the minimum in $\rho_{c}$ for our  sample is barely reached at the lowest
temperature  near $T_c$ (see Fig. 2a,  where the maximum of  $\ell \approx 82\AA\approx 2\xi$ ). 

The origin of Eq. (3) can be understood from the conventional quasiclassical
description
of the in-plane conductivity:
\begin{equation}
\sigma_{ab}=  \frac{e^2n\tau}{m},
\end{equation}
where $\tau$ is the relaxation time of the distribution function.
The phase coherence length can be expressed as
\begin{equation}
\ell=  v_F\tau_{\varphi} +\xi,
\end{equation}
where  $\tau_{\varphi }$ is the decoherence time and 
$v_F$ is the Fermi velocity. We introduced the empirical  cutoff   $\xi$ to indicate that
the phase coherence length in a crystal  
does not scale to zero with decreasing $\tau_{\varphi }$. 
The value of $\xi$ should correlate  with  elastic
mean free path, because  elastic collisions do not cause  loss of phase coherence.  Equation  (9)
reflects the ballistic (not diffusive) nature of  the motion between the phase relaxation events.
If we assume that the decoherence time and the relaxation time of the distribution
function
are proportional to each other, i.e. $\tau_{\varphi} =A\tau$ ($A=const$), Eq. (3)
follows with $q=e^2n/m v_FA$.

If, instead of Eq. (9),  $\ell$ is determined by diffusion, i.e., 
$\ell^2=D\tau_{\varphi } +\xi^2$, then, instead of Eq. (5), there is the following linear correlation
between  
$\rho_c$ and $\rho_{ab}$:  $\rho_c= \beta_c +(\xi^2/\ell_0^2)\rho_{ab}$.   
We find, however, that  the ballistic
motion of the electrons between phase relaxation collisions, namely Eqs. (3) and (9),
describes appreciably better the resistivity data of $YBa_{2}Cu_{3}O_{6.93}$.

In summary, we have analyzed  the resistivity data of $YBa_2Cu_3O_x$ ($6.35 < x <
6.93$) assuming that the interlayer  phase coherence length  is
temperature independent, equal to the spacing between
the neighboring bilayers. Under this condition, the out-of-plane transport is
non-classical and $\sigma_c$ cannot be obtained from Boltzmann-Landau 
kinetic equation. Instead, we use the relationship between conductivities given
by Eq. (1).  Then, the out-of-plane conductivity is determined by the relationship between 
the in-plane  phase coherence length and the relaxation time $\tau$. This relationship 
may be different for different systems.  We also found that $\sigma_{ab}$ of single crystals
with different  oxygen content
can be  well described by a universal two parameter
dependence on the in-plane TL
which exhibits a branching point corresponding to the metal-insulator
transition.



\section*{ACKNOWLEDGMENTS}
This research was supported by the National Science Foundation under Grant No.
DMR-9801990 at KSU and the US Department of Energy under  Contract No.
W-31-109-ENG-38 at ANL.

\end{document}